\documentstyle[aps,multicol,epsfig,epsf]{revtex}

\begin{document}

\title{Evolution of percolating force chains in compressed granular
  media} 
\author{ Ra\'{u}l Cruz Hidalgo$^1$, Christian U.
  Grosse$^2$, Ferenc Kun$^{1,3}$, Hans W. Reinhardt$^2$, and  Hans
  J.\ Herrmann$^1$}   
\address{$^1$ ICA1, University of Stuttgart, Pfaffenwaldring 27, 70569
Stuttgart, Germany.\\ 
$^2$ IWB, University of Stuttgart, Pfaffenwaldring 4, 70569
Stuttgart, Germany.\\ 
$^3$ Department of Theoretical Physics, University
of Debrecen, P.\ O.\ Box: 5, H-4010 Debrecen, Hungary. }

\date{\today} 
\maketitle 
\widetext
\begin{abstract}

The evolution of effective force chains percolating through a
compressed granular system is investigated. We performed       
experiments by compressing an ensemble of spherical particles in a 
cylindrical container monitoring the macroscopic constitutive
behavior and the acoustic signals  emitted by microscopic
rearrangements of particles. As a novel approach, we applied the
continuous damage model of fiber bundles to describe the evolution of
the array of force chains during the loading process. The model
provides a nonlinear constitutive behavior in good quantitative
agreement with the experimental results. For a system of hard
particles the model predicts a universal power law divergence of
stress when approaching a critical deformation. 
The amplitude distribution of acoustic signals was found experimentally to
follow a power law with exponent $\delta = 1.15 \pm 0.05$ which is in
a good agreement with the analytic solution of the model.  

\end{abstract}


\begin{multicols}{2}
\narrowtext

Recently, the behavior of granular materials has been extensively
studied under various conditions due to their scientific and
technological importance. 
Huge experimental and theoretical efforts have been devoted to obtain
a better understanding of the global behavior of granular media in terms of 
microscopic phenomena which occur at the level of discrete particles 
\cite{vis1,exp1,Miller,Stauffer,Roux,Rintoul96,Schwartz}.
Subjecting a confined granular packing to an uniaxial compression a
rather peculiar constitutive behavior can be observed:
for small strains a strong deviation from the linear elastic
response can be found implying that the system drastically hardens
in this regime \cite{exp1,Schwartz}. Linear elastic behavior can only
be achieved  asymptotically at larger deformations when the system
gets highly compacted. When the external load is decreased again the
system shows an irreversible increase in its effective stiffness,
furthermore, under cyclic loading hysteretic behavior is obtained.

Microscopically, inside a compressed granular packing, stresses are
transferred by the 
contact of particles. Under gradual loading conditions the particles get
slightly displaced changing their contacts and the local load supported
by them, which can be experimentally visualized using photoelastic
materials for grains. These experiments revealed that in a compressed
granular system  the stresses are transmitted along the direction of
the external load by 
force chains which can branch at the grains and form a complex network
\cite{vis1}. Particles lying between lines of the force network do not
support any load and can even be removed from the packing
without changing its mechanical properties. Increasing
the external load, more and more force lines appear
and they all undergo erratic changes until the system
reaches a saturated state when all the particles hold typically the
same load and the system behaves as a bulk material. The creation and
restructuring of percolating force chains implies relative displacements of
particles which can be followed experimentally by recording the
acoustic waves emitted, however, up to now no such experiments have been 
performed systematically. 
Theoretically, this problem has been mainly studied by means 
contact dynamics simulations using 
spherical or cylindrical particles, and cellular automata
\cite{Roux,Rintoul96,Schwartz}.   
Computer simulations also revealed the
generation and evolution of force chains in compressed granular
materials, however, the statistics of microscopic restructuring
events, the emergence of the array of force chains and their relation
to the macroscopic constitutive behavior remained unclarified. 
\begin{figure}
  \begin{center}
\epsfig{file=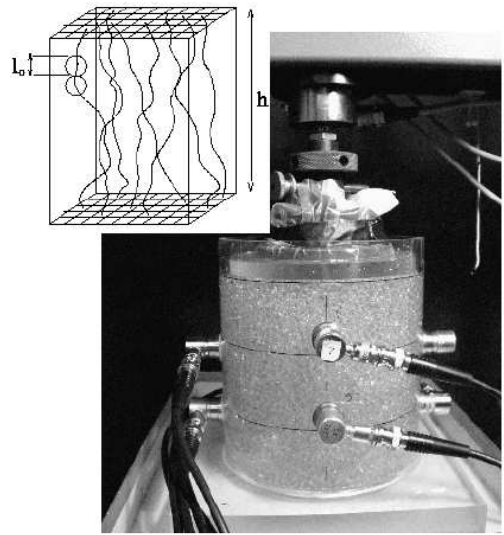,bbllx=150,bblly=300,bburx=430,bbury=645,
width=7cm,angle=0}
\vskip 0.1cm           
  \caption{Experimental set up and sketch of the array of force
    chains used in the model.}
\label{fig:setup}   
  \end{center}
\end{figure}          
In the present letter the generation and evolution of percolating
force chains is studied experimentally and theoretically in granular
packings subjected to an uniaxial external load.  We measured
experimentally the macroscopic constitutive behavior and the acoustic
signals emitted by microscopic restructuring events compressing an ensemble
of spherical glass bead confined in a cylinder. Based on the analogy of
force lines percolating through the system and fibers of a fiber
composite we propose a novel theoretical approach, namely, an 
inversion of the Continuous Damage Model (CDM) of fiber bundles
\cite{feri1,feri2} to describe the stress transmission through
granular assemblies. The model 
naturally captures the emergence and gradual hardening of force chains
and provides analytic solutions for the constitutive behavior and
acoustic activity as well. We also propose an efficient Monte Carlo
simulation technique. 

\begin{figure}
 \begin{center}
  \epsfig{file=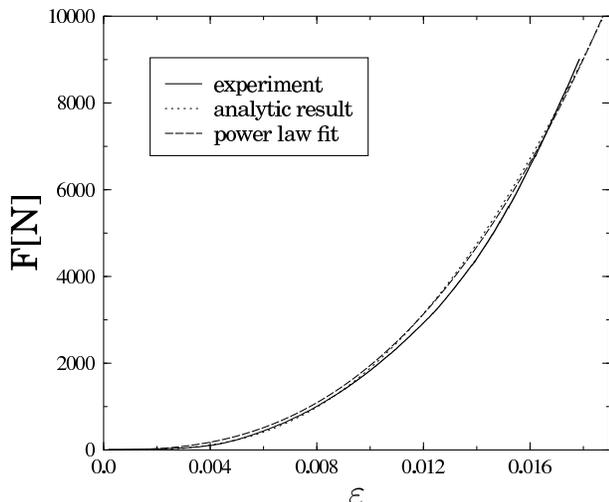,width=8cm}     
  \caption{Constitutive behavior measured experimentally. A power law
    with exponent 2.6 was fitted to the data. The comparison with
    the analytic solution of the model is also presented.} 
 \label{fig:const}
\end{center}
\end{figure} 
In our experiments a cylindrical container made of PMMA was
filled with glass beads of 5 mm diameter and water. The cylinder has a
thickness of 5 mm and a diameter of 140 mm.
A punch test was carried out applying monotonically increasing
displacements at the top level of the  
glass beads. Eight acoustic sensors were placed at 
the container wall to record the signals emitted during the compression  
of the beads, as can be seen in Fig.\ \ref{fig:setup}. 
The actual force and the displacement, measured at 
the traverse of the loading machine, and the acoustic signals were 
recorded simultaneously.   
An eight channel transient recorder was used as an analogue-digital 
converter to enable the storing of the  acoustic emission 
waveforms and a signal-based data. 
Experiments were performed under strain controlled conditions at a
fixed strain rate, {\it i.e.} moving the traverse at a constant speed of 1
mm/minute.  
 
The nonlinear elastic response of the system can be observed in Fig.\
\ref{fig:const} where the measured force $F$ is presented as a function of
strain $\varepsilon$ imposed. 
A power law of an exponent 2.6 was obtained as a best fit to the
measured data in reasonable agreement with former experiments 
of Ref.\ \cite{exp1}. 

\begin{figure}
\begin{center}
\epsfig{bbllx=140,bblly=375,bburx=480,bbury=675,
  file=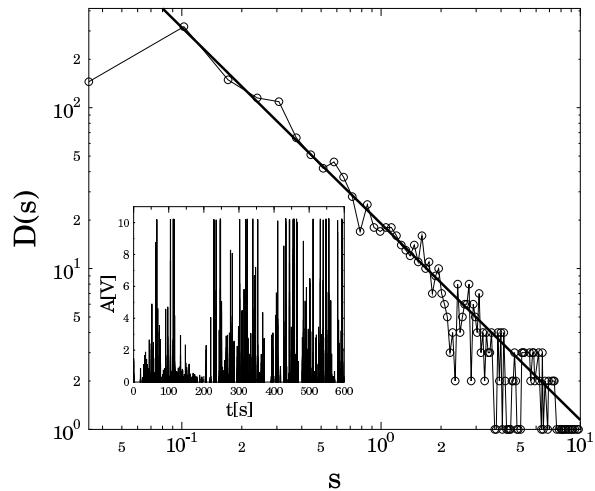, width=8.0cm} 
\caption{The statistics of acoustic signals. A power law of an
  exponent $\delta = 1.15 \pm 0.05$ was fitted to the size
  distribution of the signals of the inset.}
\label{fig:stat_exp}
\end{center}
\end{figure}

The emergence and gradual hardening of force chains is responsible on
the microscopic level for the strong non-linearity observed
macroscopically. To obtain information about microscopic processes,
the acoustic waves emitted due to sudden relative displacements of
particles were monitored.
Typically several hundred signals were recorded during the experiment.
The inset of Fig.\ \ref{fig:stat_exp}  
shows the automatically extracted peak amplitudes of the burst signals
versus time. The energy is defined as the integral of 
the acoustic emission signal amplitude following the onset time. 
The energy values of the acoustic emissions are summed up in intervals of 30 
seconds to elucidate the time dependent evaluation of acoustic emission 
activity.  More
details of acoustic emission data analysis and especially signal-based
techniques can be found in \cite{grosse1,grosse2}. 
The statistics of restructuring events is characterized by the
distribution $D(s)$ of the height of peaks $s$,
which is presented in Fig.\
\ref{fig:stat_exp} on a double logarithmic plot. It can be seen that
$D(s)$ shows a power law behavior over two orders of magnitude, the
exponent of the fitted straight line is $\delta = 1.15 \pm 0.05$.

Based on the analogy of force chains in granular packings and fibers
of fibrous materials or fiber reinforced composites, we propose a
novel theoretical approach, namely, an inverted fiber bundle model, to
describe the gradual emergence and hardening of force chains occurring
under uniaxial compression. Fiber bundles are composed of
parallel fibers of identical elastic properties but stochastically
distributed breaking thresholds. A fiber fails during the loading
process when the local load on it exceeds its breaking
threshold. Fiber failures are followed by a redistribution of load on
the remaining intact fibers according to the range of interaction in
the system. The so-called Continuous Damage Model introduced recently
\cite{feri1,feri2}, is particularly suited to model granular materials
since it captures gradual stiffness changes of elements of the model. 
In our model of compressed packings, force lines formed by particles
are represented by an array of lines organized in a square lattice
as illustrated in the inset of Fig.\ 
\ref{fig:setup}. A randomly distributed rearrangement thresholds $d$
is assigned to each line of the array from a cumulative
probability distribution $P_0(\frac{d}{d_c})$, where $d_c$ denotes the
characteristic strength of force lines.
During the compression process, 
when the local load on a line exceeds its threshold value $d$  
the line undergoes a sudden restructuring as a result of which it
becomes stiffer and straighter. The lines' stiffness increases
in a multiplicative manner, {\it i.e.} the stiffness is
multiplied by a factor $a>1$ at each restructuring so that the
constitutive equation of a single line after suffering $k$
restructurings reads as $\sigma = E_{\rm o}a^k
\varepsilon^{\alpha}$. Here $E_{\rm o}$ denotes the stiffness
characterizing single particle 
contacts, and the exponent $\alpha$ takes into account possible
non-linearities of a single contact like for Hertz law $\alpha = 1.5$.   
After each rearrangement the force chain gets a new threshold value
(annealed disorder) from a distribution of the same functional form,
but the characteristic strength $d_c$ of the distribution is increased
in a multiplicative way so that after $k$ rearrangement events the
disorder distribution takes the form
$P_k(\frac{d}{d_c})=P_0(\frac{d}{d_oq^{k}})$. The maximum value of
possible restructurings $k_{max}$ is proportional to the number of contacts and therefore to $\frac{h}{l_o}$ in the nomenclature of the inset of 
Fig. \ \ref{fig:setup}.
The ratio $\tau = a/q$
is a very important parameter of the model, it decides whether the
force chain becomes more fragile ($\tau > 1$) or more ductile ($\tau <
1$) as a result of restructuring. Essentially, $\tau$ is material
dependent and determined by the stability of
contacts. Calculations showed that the most interesting behavior of
the model can be found when $\tau = 1$, hence, in the present letter
our investigation is restricted to this case. The full complexity of
the model will be explored in a forthcoming publication
\cite{2version}. 

Experiments and discrete element simulations \cite{vis1,Schwartz} have 
revealed that the
number of effective force chains increases during the compression
process until it reaches a saturation value. To capture this effect in
our model, for the number of elements we prescribe the form
$N(\varepsilon)=N_oG(\varepsilon)$, where $N_o$ denotes the saturation
number of chains, and the profile $G(\varepsilon)$ has the property
$G(\varepsilon) \rightarrow 1$ with increasing $\varepsilon$. Hence,
the number of force lines $dN$ emerging due to an infinitesimal
deformation increment from $\varepsilon$ to $\varepsilon+d\varepsilon$
is $dN = N_og(\varepsilon)d\varepsilon$, where $g(\varepsilon) =
dG(\varepsilon)/d\varepsilon$. Following the derivation of the
constitutive behavior of the continuous damage model of fiber
bundles \cite{feri1,feri2}, the macroscopic constitutive equation of
the compressed granular system can be cast into the form
\begin{eqnarray}
  \label{eq:constit}
   \sigma(\varepsilon) &=& E_o \sum_{i=0}^{k_{max}-1} \int_{0}^{\varepsilon}a^i 
     (\varepsilon-\varepsilon^*)^\alpha g(\varepsilon^*) 
     P^{i}_{0}(\varepsilon-\varepsilon^*) \times \\ 
     && \left[1-P_0(\varepsilon-\varepsilon^*)\right] d\varepsilon^* \nonumber
     \\  
      &+&  \int_{0}^{\varepsilon}  a^{k_{max}}
      (\varepsilon-\varepsilon^*)^\alpha  g(\varepsilon^*)
      P^{k_{max}}_{0}(\varepsilon-\varepsilon^*) d\varepsilon^*. \nonumber
\end{eqnarray}
Eq.\ (\ref{eq:constit}) takes also into account that the local strain
of force lines $\varepsilon-\varepsilon^*$ is different from the
externally imposed strain value $\varepsilon$ since it also depends on
the initial strain $\varepsilon^*$. The integral is
performed over the whole loading history to take into account all the
lines generated. For explicit calculations we imposed an exponential
form $\displaystyle{N(\varepsilon) = N_o(1-e^{-\frac{\varepsilon}{\beta}})}$
for the number of chains. The best fit obtained to the experimental
data is presented in Fig.\ \ref{fig:const} where the force $F=N_o \sigma$ is
plotted against deformation $\varepsilon$. It can be seen that a good
fit was achieved with physically reasonable parameter
values. The maximum possible number  
of percolating force chains $N_o$ that can emerge in the system was
estimated as the ratio of the total area of the container $A_o$ to
the cross section of a single particle $A_p$, {\em i. e.}
$N_o=\frac{A_o}{A_p}=784$.  The value of the other parameters are 
$E_o=4600 \frac{N}{m^2}, d_o= 3 N$, $\beta=0.01$ and $a=q=1.01$. 
The value of $a$ falls close to one indicating that a single
restructuring gives rise only to a slight increase of stiffness of a
force chain. Model calculations revealed that the zero derivative at 
the starting part of the constitutive curve is due to the gradual 
creation of load bearing force chains. 
The small value of $\beta$ implies that
the generation of new force chains stops at a relatively small strain
value, and hence, the later rapid increase of $F$ as  function of
$\varepsilon$ is mainly caused by the hardening of the existing force
lines. Further information can be gained about the constitutive
behavior for larger strains by simplifying Eq.\ (\ref{eq:constit}) assuming 
a fixed number $N_o$ of force chains from the beginning
of the process. Under this assumption Eq.\ (\ref{eq:constit}) can be 
reformulated as
\begin{eqnarray}
\label{eq:constit_simp}
        \sigma(\varepsilon) &=& E_o \varepsilon 
        \left[1- P_0(\varepsilon)\right] \sum_{i=0}^{k_{max}-1} 
        a^i P_0(\varepsilon)^i \\ \nonumber
         &+& a^{k_{max}}E_o \varepsilon P_0(\varepsilon)^{k_{max}}.
\end{eqnarray}
It can be seen from Eq.\ (\ref{eq:constit_simp}) that if the maximum
number $k_{max}$ of possible restructuring events goes to infinity
the stress $\sigma$ has finite values only for  $aP_0(\varepsilon)<1$. 
In this case the summation can be performed in the first term,
while the second term tends to zero, and  
the constitutive equation takes the form
\begin{equation}
\label{eq:cons_beh_a3}
        \sigma(\varepsilon) = E_o \varepsilon 
        \left[1-P_0(\varepsilon)\right] \frac{1}{1- a
          P_0(\varepsilon)}. 
\end{equation}
It follows that the stress $\sigma$ diverges when $\varepsilon$
approaches a critical value $\varepsilon_c$, where $\varepsilon_c$
satisfies the equation
$\displaystyle{P_0(\varepsilon_c)=1/a}$. Expanding 
$P_0(\varepsilon)$ into a 
Taylor series at $\varepsilon_c$ as $P_0(\varepsilon) =
P_0(\varepsilon_c) +  p_0(\varepsilon_c)(\varepsilon_c-\varepsilon) +
\ldots $, and substituting it into Eq.\ (\ref{eq:cons_beh_a3}) the
behavior of $\sigma$ in the vicinity of $\varepsilon_c$ reads as
\begin{equation}
 \sigma(\varepsilon) \approx
 \frac{1}{ap_0(\varepsilon_c)(\varepsilon_c-\varepsilon)}  \sim
 (\varepsilon_c-\varepsilon)^{-1}.
\label{eq:exponet}
\end{equation}
It means that the stress $\sigma$ shows a power law
divergence when $\varepsilon$ approaches the critical value
$\varepsilon_c$. The value of the exponent is universal; it does not
depend on the form of disorder distribution $P_0$, while the value of
$\varepsilon_c$ depends on it. It is interesting to note
that in Ref.\ \cite{Rintoul96} the same power law divergence was found
in large scale molecular dynamic simulations of a hard sphere 
system.   

\begin{figure}
 \begin{center}  
   \epsfig{file=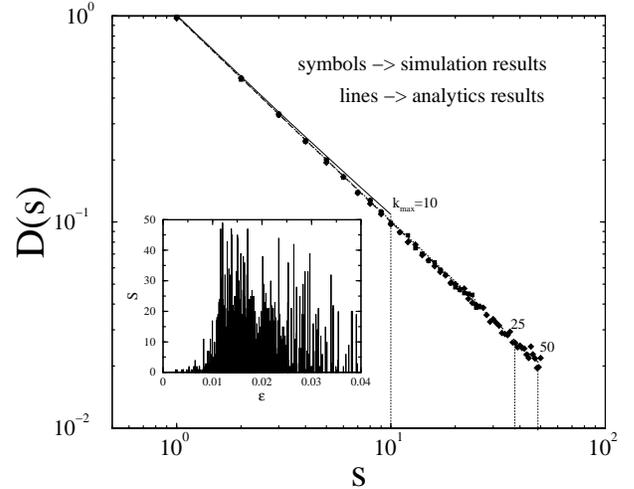,angle=-90,width=8cm}        
  \end{center} 
  \caption{Inset: local events of size $s$  from a Monte Carlo
    simulation of  strain controlled loading. 
 The statistics of events $D(s)$ is presented in a double logarithmic
 plot.}  
  \label{fig:dist_total} 
\end{figure}

When the number of force lines is fixed it is possible to obtain
analytic results also for the statistics of restructuring events. 
Restructuring occurs during the compression process
when the local load on a force line exceeds its threshold value. Since
loading is performed under strain controlled conditions,
there is no load redistribution among existing force lines, {\it i.e.}
restructuring of a force line does not affect other elements of the system.
If the new threshold value, assigned to the force line after
rearrangement, is smaller than the local load, the
force line undergoes successive restructurings until it gets
stabilized. The number of steps to reach the stable state defines the
size $s$ of the restructuring event, which is the analog of the
acoustic signals measured experimentally.
The  number of restructuring events $N_{k,s}$ of size $s$
starting in force chains which have already suffered $k$ restructurings
can be deduced as   
\begin{eqnarray}
\frac{N_{k,s}(\varepsilon)}{N_o} =
p_0(\varepsilon) P_0^{s+k-1}(\varepsilon)
\left[1-P_0(\varepsilon)\right],
\label{eq:ks}
\end{eqnarray}
for $s + k \le k_{max} - 1$, and 
\begin{eqnarray}
\frac{N_{k,s}(\varepsilon)}{N_o} = p_0(\varepsilon)
P_0^{k_{max}-1}(\varepsilon),
\label{eq:N_ki1}
\end{eqnarray} 
for $s+k=k_{max}$ (see also Ref.\ \cite{feri2}). The number of events
$D(s)$ of size $s$ can be 
determined by integrating over the entire loading history and summing
over all possible $k$ values
\begin{eqnarray} 
D(s) = \sum_{k=0}^{k_{max}-s} \int_{0}^{\varepsilon_c}  
\frac{N_{k,s}(\varepsilon)}{N_o}
d\varepsilon  
+ \int_{0}^{\varepsilon_c} 
\frac{ N_{k_{max}-s,s}(\varepsilon)}{N_o} d\varepsilon.
\label{eq:dist} 
\end{eqnarray}
Finally, substituting Eqs.\ (\ref{eq:ks},\ref{eq:N_ki1}) into Eq.\
(\ref{eq:dist}) and performing the calculations yields 
\begin{eqnarray}
  \label{eq:final}
  D(s) = s^{-1}, \qquad \mbox{where} \qquad 1\leq s \leq k_{max},
\end{eqnarray}
{\em i.e.} the distribution of microscopic restructuring events
exhibits an universal power law behavior with an exponent 1, which is
completely independent on the disorder distribution. Numerical
simulations revealed that the universal power law behavior also holds
when the gradual creation of force chains is taken into account, {\em
  i.e.} when the system is described by the full  Eq.\
(\ref{eq:constit}). The statistics of restructuring events obtained by
Monte Carlo simulations is presented in Fig.\ \ref{fig:dist_total},
where simulations were performed under the assumption 
$\displaystyle{N(\varepsilon) = N_o(1-e^{-\frac{\varepsilon}{\beta}})}$. 
The inset shows local events of different sizes that occurred during the
loading process, and their distribution is presented in the main figure. 
The power law behavior of the analytic prediction of eq.\ \ref{eq:final} is
verified. It is important to emphasize that the theoretical results on
event statistics (Fig.\ \ref{fig:dist_total}) are in a very good
quantitative agreement with the 
experimental findings (Fig.\ \ref{fig:stat_exp}). 
Note, that the functional form and the value of
the exponent of $D(s)$ in the analytic calculations is mainly the
consequence of the locality of restructurings due to the absence of
load redistribution. The excellent agreement
observed indicates that this is likely the microscopic
mechanism responsible for the power law statistics of acoustic signals
observed experimentally.

This work was supported by the project SFB381, and by the NATO grant
PST.CLG.977311. F. Kun  acknowledges financial support of the
B\'olyai J\'anos Fellowship of the Hungarian Academy of Sciences and
of the Research Contract FKFP 0118/2001.

\end{multicols}

\begin{references}

\bibitem{vis1} C.\ Liu, S.\ R.\ Nagel, D.\ A.\ Schecter, S.\ N.\ Coppersmith, 
S.\  Majumdar, O.\ Narayan, and J.\ P.\ Witten, Science {\bf 259}, 513 (1995). 
\bibitem{exp1} T.\ Travers, D.\ Bideau, A.\ Gervois, J.\ P.\ Troadec, and 
J.\ C.\ Messager, J.\ Phys.\ A {\bf 74}, 19 (1986).    
\bibitem{Miller} B.\ Miller, C.\ O'Hern, and R.\ P.\ Behringer,
Phys.\ Rev.\ Lett.\ {\bf 77}, 3110 (1996).
\bibitem{Stauffer} H.\ J.\ Herrmann, D.\ Stauffer, and S.\ Roux, 
Europhys.\ Lett.\ {\bf 3}, 265 (1987).   
\bibitem{Roux} F.\ Radjai, M.\ Jean, J.-J.\ Moreau, and 
S.\ Roux, Phys.\ Rev.\ Lett.\ {\bf 77}, 274 (1996).
\bibitem{Rintoul96}M.\ D.\ Rintoul and S.\ Torquato, Phys.\ Rev.\ Lett.\
{\bf 77}, 4198 (1996).
\bibitem{Schwartz}H.\ A.\ Makse, D.\ L.\ Johnson, and 
L.\ M.\ Schwartz, Phys.\ Rev.\ Lett.\ {\bf 84}, 4160 (2000).
\bibitem{feri1} F.\ Kun, S.\ Zapperi, and H.\ J.\ Herrmann,  
Eur.\ Phys.\ J.\ B {\bf 17}, 269 (2000).
\bibitem{feri2} R.\ C.\ Hidalgo, F.\ Kun, and H.\ J.\ Herrmann, 
Phys.\ Rev.\ E {\bf 64}, 066122 (2001). 
\bibitem{grosse1} C.\ U.\ Grosse, H.\ W.\ Reinhardt, and T.\ Dahm, 
NDT\&E Intern.\ {\bf 30}, 223 (1997).  
\bibitem{grosse2} C.\ U.\ Grosse, B.\ Weiler, H.\ W.\ Reinhardt, 
J.\ of Acoustic Emission {\bf 14}, 64 (1997).
\bibitem{2version} R.\ C.\ Hidalgo, C.\ U.\ Grosse, F.\ Kun,  
H.\ W.\ Reinhardt, and  H.\ J.\ Herrmann. unpublished (2002)

\end{references}
\end{document}